\documentstyle[tighten,aps]{revtex}

\newcommand{\be}{\begin{equation}}
\newcommand{\ee}{\end{equation}}
\newcommand{\ba}{\begin{eqnarray}}
\newcommand{\ea}{\end{eqnarray}}

\newcommand{\fr}[2]{\frac{#1}{#2}}
\newcommand{\non}{\nonumber}

\def\vec#1{{\mbox{\boldmath$#1$}}}

\newcommand{\p}{\mbox{$\vec{p}$}}
\newcommand{\q}{\mbox{$\vec{q}$}}

\newcommand{\k}{\mbox{$\vec{k}$}}

\newcommand{\r}{\mbox{$\vec{r}$}}

\newcommand{\lb}{\left (}
\newcommand{\rb}{\right )}
\newcommand{\la}{\left\langle}
\newcommand{\ra}{\right\rangle}
\newcommand{\ep}{\epsilon}
\newcommand{\vsig}{\mbox{$\vec{\sigma}$}}

\begin{document}

\title{
%
%
\[ \vspace{-2cm} \]
\noindent\hfill\hbox{\rm SLAC-PUB-8557 } \vskip 1pt
\noindent\hfill\hbox{\rm hep-ph/0008099} \vskip 10pt
%
%
${\cal O}(\alpha^3 \ln \alpha)$ corrections to positronium decay rates}

\author{Kirill Melnikov\thanks{
e-mail:  melnikov@slac.stanford.edu}}
\address{Stanford Linear Accelerator Center\\
Stanford University, Stanford, CA 94309}
\author{Alexander Yelkhovsky \thanks{
e-mail: yelkhovsky@inp.nsk.su}}
\address{Budker Institute for Nuclear Physics,\\
Novosibirsk 630090, Russia}
\maketitle

\begin{abstract}
We compute ${\cal O}(\alpha^3 \ln \alpha )$ corrections
to the decay rates of para- and orthopositronium into
two and three photons, respectively. For this calculation
we employ the nonrelativistic QED regularized dimensionally
and we explain how in this framework the logarithms
of the fine structure constant can be extracted.
\end{abstract}

\section{Introduction}

Positronium decays into two and three photons provide
an interesting  test of bound states Quantum Electrodynamics (QED).
While the parapositronium (p-Ps)
decay rate is well described by QED,  it is known that
the decay of orthopositronium (o-Ps)
into three photons is still a controversial
issue \cite{hydr2}. However, on the theoretical side there
has been a major breakthrough recently  and both decay
rates are now known with ${\cal O}(\alpha ^2)$
accuracy \cite{pPs,afs}.
The next level of precision, i.e. ${\cal O}(\alpha ^3)$
correction, is currently beyond the reach, although parts of the
${\cal O}(\alpha^3)$  correction that are enhanced by the
logarithms of the fine structure  constant, can be computed.

The knowledge of these corrections is not of significant phenomenological
importance at present since
they are much smaller than the
experimental accuracy for both p-Ps and o-Ps decays. Nevertheless,
from a theoretical viewpoint, it is an interesting problem because
at order $\alpha^3$
there are both leading ${\cal O}(\alpha^3 \ln^2 \alpha)$
and subleading ${\cal O}(\alpha^3 \ln \alpha)$ corrections.  The calculation
of the leading ${\cal O}(\alpha^3 \ln^2 \alpha)$ corrections is a fairly simple
enterprise; it has been done quite some time ago  \cite{karsh}.
Here we are interested in
subleading logarithmic corrections. In order to compute them, we utilize
the nonrelativistic QED in dimensional regularization and
we explain how in this framework the logarithmic corrections
can be computed using limited amount of information.

Before diving into the description of the calculation, let us summarize
our results. For the ${\cal O}(\alpha^3 \ln \alpha)$ corrections
to para- and orthopositronium decay rates we find:
\ba
&&\Delta \Gamma_{\rm p} = \frac{\alpha^3}{\pi} \ln \alpha
\left\{ - \frac{367}{90} + 10 \ln 2 - 2 A_{\rm p}
\right\} \Gamma_{\rm p}^{(0)},
\label{para} \\
&&
\Delta \Gamma_{\rm o} = \frac{\alpha^3}{\pi} \ln \alpha
\left\{ - \frac{229}{30} + 8 \ln 2 + \frac{A_{\rm o}}{3}
\right\} \Gamma_{\rm o}^{(0)}.
\label{ortho}
\ea
The coefficients $A_{\rm p,o}$
describe ${\cal O}(\alpha/\pi)$ corrections to the lowest order decay
widths $\Gamma_{\rm p,o}^{(0)}$. They are \cite{hb,adkins96}
\be
A_{\rm p} = \fr{\pi^2}{4} - 5,~~~~~A_{\rm o} = -10.286606(10).
\ee

Numerically, the ${\cal O}(\alpha^3 \ln \alpha)$ corrections,
Eqs.(\ref{para},\ref{ortho}), evaluate to
$7.7919~\alpha^3/\pi \ln \alpha\;\Gamma_{\rm p}^{(0)}$ for p-Ps and to
$-5.517~\alpha^3/\pi \ln \alpha\;\Gamma_{\rm o}^{(0)} $ for o-Ps.
These corrections are therefore quite comparable with the
``leading'' ${\cal O}(\alpha^3 \log^2 \alpha)$ corrections computed
in \cite{karsh}.

The paper is organized as follows. In the next Section we set up the
framework of the calculation. We then continue with detailed discussion
of how various contributions to Ps decays at ${\cal O}(\alpha^3 \ln \alpha)$
are computed. In the last Section we present our conclusions.


\section{Framework of the calculation}

Let us first discuss the framework of our calculation. We work
in nonrelativistic QED regularized dimensionally; $d=3-2\ep$
is the number of spatial dimensions and $\ep$ is the regularization
parameter. General features
of this  technique have been  described at length in our previous
paper \cite{cmy}.  Here we would like to discuss a new issue
which was not considered in \cite{cmy} -- how logarithmic in $\alpha$ corrections  can be extracted.

In order to extract logarithms of the
fine structure constant in a self-consistent way, we use the fact that
the matrix element of any operator
in dimensional regularization is a  {\it uniform } function
of the
fine structure constant. This implies that
all the dependence
on the fine structure constant can be scaled out of any dimensionally
regularized matrix element\footnote{We stress that this is the feature
of dimensional regularization and it is not valid in other
regularization schemes.}. The scaling, however,
should be done in $d$ dimensions.

In order to establish the scaling rules, we need to know
how different  quantities involved in  bound state calculations
scale with $\alpha$. To do that, we rewrite the $d$-dimensional Schr\"odinger
equation 
in ``atomic units'' familiar from the standard treatment
of hydrogen atom  in three-dimensional Quantum Mechanics.

Consider the Schr\"odinger equation for positronium in $d$ dimensions:
\be
 \left ( \frac {p^2}{m} -  \frac {c(d)\alpha}{r^{d-2}}\right )
\Psi = E \Psi,
\label{sch}
\ee
where $c(d) = \Gamma(d/2-1)/\pi^{d/2-1}$ (see \cite{cmy}).
Let us  rescale $p \to p \gamma$ and $r \to r/\gamma $ and
choose $\gamma = (m\alpha/2)^{\frac {1}{1+2\ep}}$.  Then both
the Coulomb Hamiltonian
$H = p^2/m - c(d)\alpha/r^{d-2}$ and its eigenvalue $E$ scale as
$ \gamma^2/m $. The
wave function of a bound state, being normalized to unity,
scales as $\gamma^{d/2}$.
Hence, when
expressed in atomic units, the energy of a bound state and its properly
normalized wave function depend on $d$ only.
The scaling rules above provide sufficient information to write the matrix
element of any operator in atomic units and therefore scale out
all the dependence on the fine structure constant.

There is another important point that makes the extraction of
the $\ln \alpha$ corrections possible using limited amount
of information. In order to explain it, we remind the reader
that in bound state calculations different contributions to the final
result  can be  distinguished.
In particular, there are so-called hard contributions.
The ${\cal O}(\alpha^3)$ hard corrections to the Ps decay rates
are described by the three-loop
Feynman diagrams for the process $e^+e^- \to 2(3) \gamma $,
that have to be computed exactly at the threshold.
Schematically, such diagrams generate the correction
\be
V_{\rm hard}
= \left ( \frac {\alpha}{\pi} \right )^3 \left [ \frac {t_1}{\ep^2}
+ \frac {t_2}{\ep} + t_3 \right ]
V_{\rm Born},
\label{hardcor}
\ee
( $t_{1-3}$ are some constants)
to the annihilation kernel $V_{\rm Born}$ that is responsible
for the lowest order decay rate:
\be
\Gamma_{\rm p,o}^{(0)} = \la \Psi | V_{\rm Born} | \Psi \ra \propto \Psi_0^2.
\ee
Here $\Psi_0$ stands for the positronium wave function at the origin.
The effective potential (\ref{hardcor}), in turn, generates the
correction to the decay rate,
\be
\delta_{\rm hard} \Gamma =
\la \Psi | V_{\rm hard} | \Psi \ra \propto
\left  [ \frac {t_1}{\ep^2}
+ \frac {t_2}{\ep} + t_3 \right ]
\Psi_0^2.
\label{hard}
\ee
If we  rewrite $\Psi_0$
in atomic units, Eq.(\ref{hard}) generates logarithms of the
fine structure constant. These logarithms are artificial,
since we anticipate that other, soft scale, contributions also generate
divergences  which exactly match and cancel all the divergences
in $V_{\rm hard}$.  This implies, that the logarithms
associated with the rescaling of the wave function at the
origin  get cancelled as well. Therefore,
the easiest way to avoid considering
$V_{\rm hard}$ (which is
not available at present)
is to work with  relative, rather than absolute,
corrections to the decay width.
This automatically  discourages Eq.(\ref{hard})
as the source of logarithms of the  fine structure constant
since in this case there is simply nothing to rescale.

In order to illustrate how these arguments help to compute the $\ln \alpha$
corrections, let us consider the  matrix element of a nonrelativistic
operator ${\cal O}$ that delivers ${\cal O}(\alpha^3)$ correction
to the lowest order annihilation kernel $V_{\rm Born}$.
In accordance with the above comment
we consider relative correction to the decay rate:
\be
\Delta_{\cal O} =
\fr {\delta_{\cal O} \Gamma}{
     \Gamma_{\rm p,o}^{(0)}}
     = \fr {\la \Psi |
{\cal O} | \Psi \ra }{ \la \Psi | V_{\rm Born} | \Psi \ra }.
\label{examp}
\ee
The operator ${\cal O}$  is
a function of  coordinate and
momentum operators that act on the positronium wave function. After
rescaling of all the quantities on the right hand side of Eq.(\ref{examp})
as described above, we end up with the following equation:
\be
\Delta_{\cal O} = \alpha^n
     \fr{\gamma^{3-n+l\epsilon}}{m^{3-n+k\ep}}
     \cdot
     \left. \fr {\la \Psi |
{\cal O} | \Psi \ra }{
\la \Psi | V_{\rm Born} | \Psi \ra } \right|_{\gamma = 1},
\ee
where $n=1,2$ is a power of $\alpha$ that explicitly enters
${\cal O}$ and  $l,k$ are some integer numbers. If the matrix
element is finite,
we can safely put $\ep=0$ and then the relative correction to the decay width
is $\alpha^3$ times the $\alpha$-independent ratio
and hence no logarithms of $\alpha$ appear.
Therefore, after the rescaling, the {\it only} place where
$\ln \alpha$ can come from
is the expansion of the factor $\gamma^{3-n+l\epsilon}$ in powers
of $\ep$; this implies that in order to generate the $\ln \alpha$
corrections, the nonrelativistic matrix elements should diverge and
only divergent pieces of the matrix elements have to be known to
determine logarithmic in $\alpha$ corrections to the decay rate.
Note also, that because of  the relation between
$\gamma$ and $\alpha$,
$$
\gamma = \lb \fr{m\alpha}{2} \rb^{\frac {1}{1+2\ep}},
$$
even the operators that scale as {\it integer} powers of $\gamma$ can
generate $\ln \alpha$ corrections.

Let us note that the extraction of the $\ln \alpha$ corrections
in dimensional regularization can lead to some counter-intuitive
results; for example, logarithms of the fine structure constant
are generated by the operators, that in  a more ``physical'' regularization
schemes, such as e.g. the schemes that use
either the photon mass or the momentum cut-off $\lambda$,  can only lead
to the logarithms of $m/\lambda$ but not to $\ln \alpha$. Therefore, it
appears that  individual contributions to the final result are scheme
dependent. Nevertheless, we would like to stress that once
dimensional regularization and  clear rules for extracting $\ln \alpha$
are adopted, there is no other way as to consider all possible contributions;
none of them can  be disregarded by invoking the fact that
in a different regularization scheme a particular operator cannot
generate ${\cal O}(\ln \alpha)$ correction.

The calculation of the nonrelativistic contributions that
are relevant at order ${\cal O}(\alpha^3 \ln \alpha)$ is
 described in the following Sections.
Some useful integrals, that we need in
the calculation, are summarized in Appendix.
We give intermediate  formulas  for the relative
logarithmic corrections to the
positronium decay rate expressed in
units of $(\alpha^3/\pi) \Gamma(1+\epsilon)^3(4\pi)^{3\epsilon}$.

\section{Irreducible  contribution}

This particular contribution arises as the average value of
a local operator that  comes from the Taylor expansion
of the one-loop corrections to the
annihilation kernel
in spatial momenta $\p$ of electron and positron.
For our purpose we  need
${\cal O}(\alpha p^2)$ correction to $V_{\rm Born}$.
Since this operator is constructed by Taylor expanding
in external momentum, the non-analytic dependence on $p^2$
cannot appear. Then,
from the rescaling argument  we know that only divergent
piece of the Wilson coefficient of this operator is required.
It turns out that this divergent piece can easily be
computed using rather general arguments. According to the rules
of nonrelativistic QED, we extract  Wilson coefficients of various
effective operators from the corresponding on-shell scattering amplitudes.
The key observation is that the divergence in the Wilson coefficient
of the ${\cal O}(\alpha p^2)$ operator causes the divergence
in the {\it on-shell}  annihilation process
$e^+ e^- \to n \gamma$, $n = 2,3$, and that it is
in fact a ``true'' infrared divergence that  should be
compensated by the real emission of an additional
soft photon in the same process.

In order to avoid confusion we stress that the mechanism of
cancelling this divergence
by real radiation does not apply to the bound state because of
C-parity conservation. There is no contradiction, however.
The real infrared divergences
in the bound state calculations are absent because electron and positron in positronium
are off shell.
The divergence appears only when we put them on mass
shell in order to extract the Wilson coefficient of the relevant operator.
The crucial observation is that considering the process in a
different kinematic regime (on-shell annihilation), we easily find a divergent piece of the appropriate Wilson coefficient from the known amplitude of the real soft photon emission.

Requiring that
virtual and real corrections to the on-shell annihilation  process $e^+e^- \to n \gamma $ add up to a finite quantity,
we get the following correction to the  annihilation kernel
$V_{\rm Born }$:
\be
V_{\rm irr} =  
\frac {2 \alpha}{3 \pi \ep} \frac {p^2 + p'^2}{m^2} 
V_{\rm Born }.
\label{irr}
\ee

Applying the rescaling arguments, we
end up with the following correction induced by the irreducible operator (\ref{irr}):
\be
\Delta_{\rm irr} = \frac {4}{3} \ln \alpha.
\label{irrres}
\ee

\section{``Hard loop" contributions}

These contributions arise in the second order of the
nonrelativistic perturbation theory. This means that the
corresponding  nonrelativistic operators ${\cal O}$ are
of the form $VGV'$, where $G$ is the
reduced Green function of the Coulomb Hamiltonian
from Eq.(\ref{sch}) and $V,V'$ are some
local operators with  one of the two originating
from a hard one-loop correction.

There are two sources of the ``hard loop" contributions.
The first one  is the
one-loop renormalization of the annihilation kernel:
\be
V_{\rm p,o} =
\frac {\alpha}{\pi} A_{\rm p,o}
V_{\rm Born }.
\label{ha}
\ee
The corresponding ${\cal O}(\alpha^3 \ln \alpha)$
correction to the decay rate is then easily
related to the  ${\cal O}(\alpha^2 \ln \alpha)$ correction computed
in \cite{ky,cl}. We find:
\be
\Delta_{\rm p,o} = \lb \fr{7}{6} \vec{S}^2 - 2 \rb
A_{\rm p,o} \ln \alpha.
\label{kernres}
\ee

The second ``hard loop'' contribution corresponds to the
``hard'' piece in ${\cal O}(m \alpha^5)$ effective potential.
It reads \cite{ps}:
\be
V_{\rm hl} = -\frac {\alpha^2}{3 m^2}
\frac {\Gamma(1+\ep)}{(4\pi)^{-\ep}} \left [
 \frac {1}{\ep} + \frac {39}{5} - 12\ln 2
+ \vec {S}^2 \left ( \frac {32}{3}+6\ln 2 \right )
- 2 \ln m
 \right ] \delta (\r),
\label{1lh}
\ee
where $\vec {S}$ is the operator of the total spin. There is no problem
with defining total spin operator here, since it multiplies explicitly
finite quantity.

The ${\cal O}(\alpha^3)$ correction to the decay rate
generated by the potential from Eq.(\ref{1lh}) then reads:
\be
\delta_{\rm hl} \Gamma =
2 \la \Psi | V_{\rm Born } G V_{\rm hl}
| \Psi \ra,
\ee
and is proportional
to the Green function at the origin, $G(0,0)$. All necessary results for
this Green function can be found in \cite{cmy}. Finally, we obtain:
\be
\Delta_{\rm hl}
=\frac {\ln^2 \alpha}{3}
+ \left ( -\frac {1}{6\ep} + 2\ln 2 - \frac {59}{30}
-\vec {S}^2 \left ( \frac {16}{9} + \ln 2 \right )
+ \ln m \right )
\ln \alpha.
\label{ehl}
\ee
Let us note  that since there is an explicitly divergent term in
Eq.(\ref{ehl}), one may wonder whether or not the difference
in the energy $E$  of the bound state
in $d$ and three dimensions should be
taken into account. We have checked that the cancellation of all
divergent  terms in the final result for the
${\cal O}(\alpha^3 \ln \alpha)$ correction
to the decay rate occurs  for  arbitrary $E$ and
for this reason we use three dimensional expression for this (rescaled)
quantity to present individual contributions as well.

\section{Seagull contribution}

The correction to the decay rate,
\be
\delta_{\rm s} \Gamma =  2\la \Psi | V_{\rm Born } G V_{\rm s}
| \Psi \ra,
\label{s-init}
\ee
is induced by the double seagull effective potential $V_{\rm s}$,
\be
V_{\rm s} (\q) = -\fr{ \pi^2\alpha^2 }{ m^2 }
\int \fr{ d^d k }{ (2\pi)^d }
\fr{ {\cal P}_{ij}(\k) {\cal P}_{ij}(\k') }{k k'(k+k')},
\ee
where $\k'=\q-\k$ and ${\cal P}_{ij}(\k) = \delta_{ij} - k_i k_j/k^2$.

To facilitate the calculation of this potential, we introduce
auxiliary integration variable $k_0$ and rewrite $V_{\rm s}(\q)$ as follows:
\be
V_{\rm s} (\q) = -\fr{ \alpha^2 }{ m^2 }
\int_0^\infty \fr{dk_0}{2\pi}
\int \fr{ d^d k }{ (2\pi)^d }
T_{ij}(\k) T_{ij}(\q-\k),
\ee
with
\be
T_{ij}(\k) = \fr{4\pi{\cal P}_{ij}(\k)}{k^2+k_0^2}.
\ee

It is convenient to consider  Fourier transform of $V_{\rm s}(\q)$:
\be
V_{\rm s} (\r) = -\fr{ \alpha^2 }{ m^2 }
\int_0^\infty \fr{dk_0}{2\pi}
T_{ij}(\r)T_{ij}(\r),
\label{vr}
\ee
where $T_{ij}(\r)$ stands for
\be
T_{ij}(\r) = \delta_{ij}
\int \frac {d^d k }{ (2\pi)^d }
e^{i\k\r}  \frac {4\pi}{k^2+k_0^2}
+ \frac { 1 }{ k_0^2 } \partial_i \partial_j
\int \frac {d^d k }{ (2\pi)^d }
e^{i\k\r}
\lb \frac {4\pi}{k^2} - \frac {4\pi}{k^2+k_0^2}
\rb.
\label{tij}
\ee

To compute $T_{ij}(\r)$ we use
\be
\int \fr{ d^d k }{ (2\pi)^d }
e^{i\k\r} \fr{4\pi}{k^2+k_0^2} =
2 \lb \fr{k_0}{2\pi r} \rb^{\fr{d}{2}-1}
K_{\fr{d}{2}-1} (k_0 r),
\label{yukawa}
\ee
where $K_\nu (x)$ is the modified Bessel function
of the second kind.

Inserting (\ref{yukawa}) to (\ref{tij}), and integrating over $k_0$
in Eq.(\ref{vr}), we obtain:
\be
V_{\rm s} (\r) = - \fr{\alpha^2 r^{-3+4\epsilon}
                      }{ 2\pi m^2 } \frac {\Gamma(1-\ep)^2}{(4\pi)^{-2\ep}}
\left[ 1 - \epsilon
\fr{17 - 8\ln 2}{3}
        + {\cal O}(\epsilon^2) \right].
\ee

We then rescale  the relative correction to the
decay rate and obtain:
\be
\frac {\delta_{\rm s}\Gamma}{\Gamma_{\rm p,o}^{(0)}} = - \fr{\alpha^2 \gamma^{1-4\ep}}{2\pi m}
 \frac {\Gamma(1-\ep)^2}{(4\pi)^{-2\ep}}
\left[ 1 - \epsilon
\fr{17 - 8\ln 2}{3}
        + {\cal O}(\epsilon^2) \right]
\int d^d r ~r^{-3+4\epsilon} G(\r,0)
\fr{ \Psi(\r) }{ \Psi_0 }.
\label{dgseag}
\ee

As we explained previously, we need
only a divergent part of the integral in Eq.(\ref{dgseag}).
Since in three dimensions the Green function $G(\r,0)$
behaves as ${\cal O}(r^{-1})$
for small values of $r$, we may expand the wave function in series
around $r=0$ and keep only two first terms in such an expansion:
\be
\fr{ \Psi(\r) }{ \Psi_0 } = 1 -
\fr{c(d)r^{4-d}}{4-d} +
{\cal O}(r^{8-2d}).
\label{psiexp}
\ee
The Green function $G$ can be written as a sum of three
pieces $G = G_0 + G_1 + G_{\rm multi}$ according to the number of
Coulomb interactions\footnote{$G_0$ is the free Green function,
$G_1$ is the single  Coulomb correction to $G_0$,
and $G_{\rm multi}$ accounts for two and more Coulomb interactions.}.
In three dimensions,
$G_0 \sim r^{-1}$, $G_1 \sim \ln r$ and $G_{\rm multi} \sim r^0$
as $r \to 0$. Therefore, the first term from the expansion
Eq.(\ref{psiexp}) is sufficient to extract the  singularities
caused by the  contributions of $G_1$ and  $G_{\rm multi}$.
For $G_{\rm multi}$ we derive:
\be
\int d^d r ~r^{-3+4\epsilon} G_{\rm multi}(\r,0)
=
4\pi \int_0^{\sim 1} dr ~r^{-1+2\epsilon} G_{\rm multi}(0,0) + {\cal O}(1) =
- \fr{3}{\epsilon} + {\cal O}(1).
\ee

In order to analyze $G_{0,1}$ contributions, it is convenient
to switch to the momentum space. We obtain:
\ba
&&\int d^d r ~r^{-3+4\epsilon} G_1(\r,0)
=
\fr{4^\epsilon \pi^{3/2-\epsilon}\Gamma(\epsilon)}{
\Gamma(3/2-2\epsilon)}
\int \fr{d^d p}{(2\pi)^d}
\fr{G_1(\p)}{p^{2\epsilon}},
\label{g0rp} \\
&&\int d^d r ~r^{-3+4\epsilon} G_0(\r,0)
\lb 1 - \fr{c(d)r^{1+2\ep}}{1+2\ep} \rb
=
\fr{2 \pi}{\epsilon}
\int \fr{d^d p}{(2\pi)^d}
\fr{G_0(\p)}{p^{2\epsilon}} \lb
1 - \fr{\pi \epsilon}{p^{1+2\ep}} \rb + {\cal O}(1).
\label{g1rp}
\ea
The integrals in Eqs.(\ref{g0rp},\ref{g1rp})
can be expressed through the integrals listed in Appendix:
\ba
\int \fr{d^d p}{(2\pi)^d}
\fr{G_1(\p)}{p^{2\epsilon}} &=&
- 16\pi \left[ I_2(1+\ep,1,1) + I_1(1,1,1)
- I_2(1,1,1) \right] +{\cal O}(\ep),
\\
\int \fr{d^d p}{(2\pi)^d}
\fr{G_0(\p)}{p^{2z}} &=& - 2 I_0(1,z).
\ea

Substituting all the relevant expressions into Eq.(\ref{dgseag}) and
expanding in $\ep$, we arrive at
the final result for the seagull contribution:
\be
\Delta_{\rm s}
=\frac {3 \ln^2 \alpha }{2}
- \left ( \frac {1}{2\ep} + \frac {4 \ln 2}{3}
+ \frac {5}{3} - 3\ln m \right )
\ln \alpha.
\label{seagres}
\ee

\section{Retardation}

In this Section we  discuss the retardation effect,
caused by the exchange of a photon
with a typical momentum
of the order of the inverse Bohr radius $k \sim m \alpha$
between  electron and positron in the bound state.
The corresponding formula reads:
\ba
\delta_{\rm ret} \Gamma &=&
2 \la \Psi \left| V_{\rm Born} G
\int \frac {d^d k}{(2\pi)^d}
j_i^e~e^{i\k\r_e}
\frac {4\pi \alpha}{2k}
\frac {{\cal P}_{ij}(\k)}{k+H-E}~
j_j^{p} e^{-i\k\r_p} \right| \Psi \ra
\non \\
&& - \la \Psi \left| V_{\rm Born}
\right| \Psi \ra \la \Psi \left|
\int \frac {d^d k}{(2\pi)^d}
j_i^e~e^{i\k\r_e}
\frac {4\pi \alpha}{2k}
\frac {{\cal P}_{ij}(\k)}{(k+H-E)^2}~
j_j^{p} e^{-i\k\r_p} \right| \Psi \ra
 + {\rm H.c.},
\label{bsret}
\ea
where $\vec{j}^e = \p^e/m - [\vsig\k,\vsig]/(4m)$ and
$\vec{j}^p = \p^p/m +  [\vsig'\k,\vsig'] /(4m)$
are electron and positron currents, respectively.

The appearance of two matrix elements in Eq.(\ref{bsret}) is related
to the fact that we need the {\it second} order correction
to the wave function of a bound state. In contrast to the
first order, at second
order of perturbation theory one should be careful to maintain
the normalization of the wave function. This is the reason why
the second term in Eq.(\ref{bsret}) appears.
The general formula for the second order
correction to the wave function can be found in \cite{qm}.

When writing   Eq.(\ref{bsret}), we have assumed that
a magnetic photon with the momentum $\k$
is emitted by the electron at the point $\r_e$ and absorbed
by the positron at the point $\r_p$. Using the fact that
$H-E \sim m \alpha^2$ and $k \sim m\alpha$ for the retardation effects,
we expand the r.h.s. of Eq.(\ref{bsret}) in powers of $(H-E)/k$ and obtain:
\be
\delta_{\rm ret} \Gamma = 2 \la \Psi |
V_{\rm Born} G V_{\rm ret} |\Psi \ra
- \la \Psi | V_{\rm Born} |\Psi \ra
 \la \Psi | V_{\rm ret}' |\Psi \ra,
\label{retpot}
\ee
where the ``retardation potential'' $V_{\rm ret}$ reads:
\be
V_{\rm ret} = -\int \frac {d^d k}{(2\pi)^d}
\frac {4\pi \alpha {\cal P}_{ij}(\k)}{2k^3}
j_i^e~e^{i\k\r_e} (H-E)
j_j^{p} e^{-i\k\r_p} + {\rm H.c.},
\label{vret}
\ee
and $V_{\rm ret}' = \partial V_{\rm ret}/\partial E$.
Let us illustrate how one deals with such expressions using the
 spin-dependent parts of the currents as an example.
The corresponding contribution to $V_{\rm ret}$ then reads:
\be
V_{\rm ret}^{\rm spin} =
\frac {\pi \alpha}{16m^2}
\int \frac {d^d k}{(2\pi)^d}
\frac {[\vsig \k,\sigma_i]
[\vsig' \k,\sigma'_i]}{k^3}
\left \{ (H-E) e^{i\k\r} + e^{i\k\r}(H-E) \right \} + {\rm H.c.},
\label{vretspin}
\ee
where the relative coordinate $\r = \r_e - \r_p$ has been introduced.
It is now easy to see that if we insert this result
 into Eq.(\ref{retpot})
and use  the Schr\"odinger equation both for the wave function $\Psi$
and for the reduced Green function $G$, we  obtain a finite correction
to the decay rate. As we explained in the Introduction,
finite contributions cannot generate logarithms of the fine structure
constant.

The analysis of the spin-independent contribution to $V_{\rm ret}$
is only slightly more cumbersome.
We again pull out $H-E$ both to the right and to the left by commuting
it with either electron or positron current and obtain the following
expression for the retardation potential:
$$
V_{\rm ret} = -\frac {\pi \alpha}{m^2}
\int \frac {d^d k}{(2\pi)^d}
\frac { {\cal P}_{ij} (\k)}{k^3}
\lb \left \{
H-E,e^{i\k\r}p_i p_j \right \}
 + [p_i,[H,p_j]]e^{i\k\r} \rb
 + {\rm H.c.},
$$
where  $\{ H-E,e^{i\k\r}p_i p_j  \}$
denotes the anticommutator of the two operators.
Using this expression for the retardation potential in Eq.(\ref{retpot})
and applying equations of motion, we arrive at the
following correction to the decay rate:
\be
\delta_{\rm ret} \Gamma = -
\fr{ 4\pi\alpha }{ m^2 }
\la \Psi | V_{\rm Born} G
\left[ U_{ij}\left[C,p_i\right],
        p_j \right] | \Psi \ra,
\label{retpsi}
\ee
where $C = -c(d)\alpha/r^{d-2}$ is the Coulomb potential, and $U_{ij}(\r)$ is defined as
\be
U_{ij}(\r)  = \int \frac {d^d k}{(2\pi)^d}
\frac {{\cal P}_{ij}(\k) }{k^3} e^{i\k \r}
= \fr{  \Gamma (-\epsilon) r^{2\epsilon} }{
         6 \pi^{2-\epsilon} }
         \lb \delta_{ij} - \epsilon n_i n_j\rb.
\ee

Performing  rescaling and computing
the double commutator in Eq.(\ref{retpsi}),
we get the result for the relative correction to the decay rate,
\be
\frac {\delta_{\rm ret}\Gamma}{\Gamma_{\rm p,o}^{(0)}} = - \fr{8\alpha \gamma^{2-2\ep}}{3 m^2}
 \frac {\Gamma(2-\ep)\Gamma(3/2-\ep)}{\pi^{3/2-\ep}}
\int d^d r ~r^{-3+4\epsilon} G(\r,0)
\fr{ \Psi(\r) }{ \Psi_0 },
\ee
which is very similar to the correction induced by the
seagull potential, Eq.(\ref{dgseag}). We can, therefore, borrow
much of the analysis from the previous Section.
We finally obtain:
\be
\Delta_{\rm ret} =
2\ln^2 \alpha
- \left (   \frac {2}{3\ep} - \frac {4\ln 2}{3}  + 4
- 4 \ln m
\right ) \ln \alpha.
\label{retres}
\ee

\section{Ultrasoft contribution}

By definition, the ultrasoft contribution is due to the  photons
with energy and momentum of order $m \alpha^2$. Such soft photons
cannot resolve the structure of the bound state and for this
reason they interact directly with  the positronium as a whole.
Since the positronium is chargeless, the interaction is necessarily
of the dipole nature.

A general formula for the correction to the
decay rate caused by  ultrasoft contributions is:
\ba
\delta_{\rm us} \Gamma &=& 2
\la \Psi \left| V_{\rm Born}\; G \;
\fr{2p_i}{m}
\int \frac {d^d k}{(2\pi)^d}
\frac {4\pi \alpha}{2k}
\frac {{\cal P}_{ij}(\k)}{E-H-k}~
\fr{2p_j}{m} \right| \Psi \ra
\non \\
&& - \la \Psi \left| V_{\rm Born}
\right| \Psi \ra
\la \Psi \left|
\fr{2p_i}{m}
\int \frac {d^d k}{(2\pi)^d}
\frac {4\pi \alpha}{2k}
\frac {{\cal P}_{ij}(\k)}{(E-H-k)^2}~
\fr{2p_j}{m} \right| \Psi \ra.
\label{bsus}
\ea
After integration over directions of $\k$ and performing
the rescaling we get:
\be
\fr{\delta_{\rm us} \Gamma}{\Gamma_{\rm p,o}^{(0)}} =
\fr{ 4^{2-\ep}\pi\alpha \gamma^{2-4\ep} }{m^{2-2\ep}}
\fr{\Omega_d}{(2\pi)^d} \fr{d-1}{d}
\lb
\la 0 \left| G \;
\p
\int_0^\infty \frac {dk ~k^{-2\ep} }{k+H-E}~
[H,\p] \right| \tilde{\Psi} \ra
- \la \Psi \left|
\p
\int_0^\infty \fr{dk ~k^{1-2\ep} }{(k+H-E)^2}~
\p \right| \Psi \ra \rb.
\label{relus}
\ee
Both matrix elements in Eq.(\ref{relus}) must be computed  in
atomic units; also $\tilde{\Psi} = \Psi/\Psi_0$,
and $\Omega_d = 2\pi^{3/2-\ep}/\Gamma(3/2-\ep)$ is the $d$-dimensional
angular volume.

The second matrix element in Eq.(\ref{relus}) is easy to
compute\footnote{Recall that only the terms that are singular
for $\ep \to 0$ are needed.}:
\be
- \la \Psi \left| \p
\int_0^\infty \fr{dk ~k^{1-2\ep} }{(k+H-E)^2}~
\p \right| \Psi \ra =
- \fr{\la \Psi \left| \p
 (H-E)^{-2\ep} \p \right| \Psi \ra}{2\ep}
= - \fr{1}{2\ep}.
\ee

Then, consider the first matrix element from Eq.(\ref{relus}).
It is convenient to write it as a sum:
\be
\la 0 \left| G \; \p
\int_0^\infty \frac {dk ~k^{-2\ep} }{k+H-E}~
[H,\p] \right| \tilde{\Psi} \ra =
M_0 + M_1,
\ee
where the two terms correspond to the number of  Coulomb interactions
between the moments of emission and absorption of
the ultrasoft photon\footnote{If two or more Coulomb photons are
exchanged, the  resulting matrix element becomes finite and, in accordance
with the argument given in Section II, it cannot generate
${\cal O}(\ln \alpha)$ correction.}.

Let us  consider $M_0$. In this case, integrating over $k$ we obtain:
\be
M_0 = \fr{1}{2\ep}
\la 0 \left| G \; \lb H_0-E \rb^{-2\ep}
\p ~[H,\p] \right| \tilde{\Psi} \ra,
\ee
where $H_0 = p^2/2$ is the free Hamiltonian in atomic units. Using
\be
\p ~[H,\p] = [\p,[H,\p]] + [H,\p]~\p =
4\pi\delta(\r) + [H,\p]~\p,
\ee
we represent $M_0$ as the sum of two terms,
\be
M_0 = M_\delta + M_\psi,
\label{us0}
\ee
where
\ba
M_\delta &=& \fr{4\pi}{2\ep}
\la 0 \left| G  \lb H_0-E \rb^{-2\ep}
\right| 0 \ra, \\
M_\psi &=& \fr{1}{2\ep}
\la 0 \left| G  \lb H_0-E \rb^{-2\ep}
[H,\p]\;\p \right| \tilde{\Psi} \ra.
\label{mpsi}
\ea

We now analyze the two terms in Eq.(\ref{us0}) separately.
In order to compute $M_\delta$ we use the fact that only two first terms
in the expansion of the Green function in the number of Coulomb exchanges, $G_0$ and $G_1$,  diverge at zero spatial separation.
Since $G_{\rm multi}(0,0)$ is finite, we can compute it for
$d=3$. On the other hand, both $G_0$ and $G_1$
are known explicitly (see e.g. \cite{cmy}) and hence the corresponding
integrals can easily be computed. Using expressions for integrals
summarized in Appendix, we arrive at the following expression for
$M_\delta$:
\be
M_\delta = - \fr{ 2^{2+2\ep} \pi }{ \ep }
I_0(1+2\ep,0)
- \fr{ 2^{5+2\ep} \pi^2 }{ \ep }
I_1(1+2\ep,1,1) - \fr{3}{\ep}.
\ee

Consider the second matrix element $M_\psi$.
In this case, it is sufficient to separate
$G = G_0 + (G-G_0)$. Since there is an overall
factor $\ep^{-1}$ in Eq.(\ref{mpsi}),
the finite matrix element that contains $G-G_0$ can be
calculated in three dimensions. Using
the expression for $G-G_0$  from \cite{cmy}, we derive:
\be
M_\psi^{\rm f} = \fr{1}{2\ep}
\la 0 \left| (G - G_0)
[H,\p]\;\p \right| \tilde{\Psi} \ra =
\fr{2}{\ep}.
\ee

The matrix element containing $G_0$
can be rewritten  in momentum space as
\be
M_\psi^{\rm i} =
\fr{1}{2\ep}
\la 0 \left| G_0 \lb H_0-E \rb^{-2\ep}
[H,\p]\;\p \right| \tilde{\Psi} \ra =
- \fr{1}{2\ep}
\int \frac {d^d p' d^d p}{(2\pi)^{2d}}
\lb \fr{2}{p'^2+1} \rb^{1+2\ep}
\fr { 4\pi(\p'-\p)\p}{(\p'-\p)^2} ~
\frac {\Psi(\p)}{\Psi_0},
\ee
where $\Psi(\p)$ is the Fourier transform of $\Psi(\r)$.
Since $M_\psi^{\rm i}$ has an overall divergence and hence we need the
integral up to a constant only, we can use the three dimensional expression
for the wave function $\Psi(\p)$:
\be
M_\psi^{\rm i} = - \fr{4\pi^2}{\ep}
\int \frac {d^d p' d^d p}{(2\pi)^{2d}}
\lb \fr{2}{p'^2+1} \rb^{1+2\ep}
\fr {(\p'-\p)\p}{(\p'-\p)^2} ~
\lb \fr{2}{p^2+1} \rb^2.
\ee
We now rewrite the scalar product in the numerator as
a linear combination of denominators
and obtain our final expression for $M_\psi = M_\psi^{\rm f} + M_\psi^{\rm i}$:
\be
M_\psi =
- \fr{4^{2+\ep} \pi^2}{\ep}
\left[ I_1(2\ep,1,2) - I_1(1+2\ep,1,1) -
I_1(1+2\ep,0,2) \right]
+ \fr{2}{\ep}.
\ee

Consider next the matrix element $M_1$. We start with the following expression:
\be
M_1 = - 
\la 0 \left| G \p
\int_0^\infty dk k^{-2\ep} \fr{1}{k+H_0-E} C
\fr{1}{k+H_0-E}[H,\p] \right| \tilde{\Psi} \ra.
\ee
Simple power counting
shows that we can safely take $\tilde{\Psi}$ at the origin,
$\tilde{\Psi}\to 1$, and also replace the Green function $G$
by its high-momentum asymptotics, $G(p) \to - 2/p^2$.
We then obtain symmetric and uniform expression
\be
M_1 = - 8\pi \int_0^\infty dk k^{-2\ep}
\int \frac {d^d p' d^d p}{(2\pi)^{2d}}
\fr{2}{p'^2(p'^2+1+2k)}\fr{4\pi~\p'\p}{(\p'-\p)^2}
\fr{2}{p^2(p^2+1+2k)}.
\ee
If we now rescale
both $\p$ and $\p'$
as $\p \to \sqrt{2k+1}~\p$, the integration over
$k$ factorizes and we obtain:
\be
M_1 = - \fr{8\pi^2}{\ep}
\left[ 2I_2(1,1,1) - 2I_1(1,1,1) - I_0(1,1)^2 \right].
\ee

Finally, using explicit expressions for the integrals from Appendix, we arrive at our final result for the ultrasoft correction to the decay rate:
\be
\Delta_{\rm us} =
-\frac {16\ln \alpha^2}{3}
+ \left ( \frac {4}{3 \ep}
+ 8 \ln 2
+\frac {20}{9} - 8\ln m \right ) \ln \alpha.
\label{us}
\ee

\section{Conclusions}

The sum of all the contributions from
Eqs.(\ref{irrres},\ref{kernres},\ref{ehl},\ref{seagres},\ref{retres},\ref{us})
gives the final result for the ${\cal O}(\alpha^3 \ln \alpha)$
corrections to  Ps decay rate:
\ba
&&\frac {\Delta \Gamma_{\rm p}}{\Gamma_{\rm p}^{(0)}}
= \frac{\alpha^3}{\pi} \left [ - \frac {3}{2} \ln^2 \alpha + \ln \alpha
\left\{ - \frac{367}{90} + 10 \ln 2 - 2 A_{\rm p}
\right\} \right ]=\frac {\alpha^3}{\pi} \left [
 - \frac {3}{2} \ln^2 \alpha +7.919 \ln \alpha \right ],
\label{parar} \\
&&
\frac {\Delta \Gamma_{\rm o}}{\Gamma_{\rm o}^{(0)}}
= \frac{\alpha^3}{\pi} \left [
 - \frac {3}{2} \ln^2 \alpha +
\ln \alpha
\left\{ - \frac{229}{30} + 8 \ln 2 + \frac{A_{\rm o}}{3}
\right\}  \right ]
=
\frac {\alpha^3}{\pi} \left [
 - \frac {3}{2} \ln^2 \alpha -5.517 \ln \alpha \right ].
\label{orthor}
\ea
Numerically,
these corrections cause a negligible change in the theoretical
prediction for  p-Ps and o-Ps lifetimes at the
current level of precision.  It is interesting to note, however,
that the magnitude  of the leading ${\cal O}(\alpha^3 \ln^2 \alpha)$ and
the subleading ${\cal O}(\alpha^3 \ln \alpha)$ corrections
is comparable; in case of o-Ps they almost cancel
each other.

Our results Eqs.(\ref{parar},\ref{orthor}) are in agreement
with two recent calculations of ${\cal O}(\alpha^3 \ln \alpha)$
corrections \cite{hill,penin}.  In Ref.\cite{hill} the result
for ${\cal O}(\alpha^3 \ln \alpha)$ correction to o-Ps decay
rate has been obtained numerically, where as in Ref.\cite{penin}
analytical methods similar to ours have been employed. We believe
that the achieved agreement between three independent calculations
ensures that the results, Eqs.(\ref{parar},\ref{orthor}), are correct.

As we mentioned, the ${\cal O}(\alpha^3 \ln \alpha)$ correction
to Ps decay rates at present is not very interesting phenomenologically.
A more important question, which we think we fully addressed in this
paper, is how the logarithms of the fine structure
constant can be efficiently extracted in the bound state calculation
when the dimensional regularization is used to regulate the nonrelativistic
dynamics. It is true that dimensional regularization offers many technical
advantages in the calculation. This does not go without a price, however,
since one has to be extremely careful in defining basic objects of the
nonrelativistic theory, e.g. the wave functions and energies. If this
is not done, one is left guessing whether or not is the calculation correct.

Our key observation, which we think cures such problems and makes our
calculation unambiguous,  is the fact that the matrix elements in
$d$  dimensions
are  the uniform functions of the fine
structure constant, and that the corresponding
power of $\alpha$ can be determined by  expressing the matrix elements
in ``$d$-dimensional'' atomic units. We think that these
arguments have not been spelled out before in the literature on one hand, and
that they are necessary to  make a convincing case, on the other.

Finally, let us note that the  technique discussed in this
paper can obviously be used in other bound state QED problems,
as well as for the heavy quarkonium states in QCD.

\section{Acknowledgments}
This work was supported in part by DOE under grant number
DE-AC03-76SF00515, and by the Russian  Foundation for
Basic Research grant 00-02-17646. We are grateful to the
BNL High Energy Theory Group, where part of this
work  has been done, for the hospitality extended to us.

\section*{Appendix}

Definition of the integrals that were used in the derivation.
\be
I_0(a,b) = \int \frac {d^d p}{(2\pi)^d}
\lb \fr{1}{p^2+1} \rb^a \lb \fr{1}{p^2}\rb^b
 = (4 \pi)^{-d/2}
 \frac {\Gamma(a+b-d/2)\Gamma(d/2-b)}{\Gamma(a)\Gamma(d/2)},
\ee
\ba
I_1(a,b,c) &&
 =
\int \frac {d^d p'~d^d p}{(2\pi)^{2d}}
\lb \fr{1}{p'^2+1} \rb^a
\lb \fr{1}{(\p'-\p)^2} \rb^b
\lb \fr{1}{p^2+1} \rb^c =
 (4\pi)^{-d}
\nonumber \\&& \times
\frac{\Gamma(a+b+c-d)\Gamma(a+b-d/2)
\Gamma(b+c-d/2) \Gamma(d/2-b)}
 {\Gamma(a) \Gamma(c) \Gamma(d/2)
 \Gamma(a+2b+c-d)},
\ea
\ba
I_2(a,b,c)&& =
\int \frac {d^d p'~d^d p}{(2\pi)^{2d}}
\lb \fr{1}{p'^2} \rb^a
\lb \fr{1}{(\p'-\p)^2} \rb^b
\lb \fr{1}{p^2+1} \rb^c =
(4 \pi)^{-d}
\nonumber \\
&& \times \frac {\Gamma(a+b+c-d) \Gamma(a+b-d/2)\Gamma(d/2-a)
 \Gamma(d/2-b)}{\Gamma(a)\Gamma(b)\Gamma(c) \Gamma(d/2)}
\ea

\end{document}